\documentstyle[manuscript,prb,aps]{revtex}
\begin{document}
\title{A Study of Quantum Many-Body Systems With
Nearest And Next-to-Nearest Neighbour Long-Range Interactions}
\author{Meripeni Ezung$^{1}$\thanks{panisprs@uohyd.ernet.in}, N.
Gurappa$^2$\thanks{gurappa@ipno.in2p3.fr}, Avinash
Khare$^3$\thanks{khare@iopb.res.in} and Prasanta K.
Panigrahi$^1$\thanks{panisp@uohyd.ernet.in}}
\address{1. School of Physics, University of Hyderabad, Hyderabad 500 046,
India\\
2. Laboratoire de Physique Theorique et Modeles Statistiques, Bat.
100,
Universite Paris-Sud, 91405, Orsay, France\\
3. Institute of Physics, Sachivalaya Marg, Bhubaneswar 751 005,
India}
\maketitle
\begin{abstract}
The scattering and bound states of the many-body systems, related
to the short-range Dyson model, are studied. First, we show that
the scattering states can be realized as coherent states and the
scattering Hamiltonian can be connected to a free system. Unlike
the closely related Calogero-Moser model, only a part of the
partial waves acquire energy independent phase shifts, after
scattering. The cause of the same is traced to the reduction in
the degeneracies. The bound state Hamiltonian for the full-line
problem is also studied and the relationship of its Hilbert space
with that of the decoupled oscillators elucidated. Finally, we
analyze the related models on circle and construct a part of the
excitation spectrum through symmetry arguments.
\end{abstract}
\draft


\newpage
\section{introduction}
Exactly solvable and quantum integrable many-body systems, with
long-range interactions, are one of the most active fields of
current research. The Calogero-Sutherland model
(CSM)\cite{cal,sut1,sut2}, the Sutherland model (SM)\cite{suthd}
and their variants are the most prominent examples among such
systems \cite{olpe}. These models have found application in
various branches of physics \cite{al,book,apo,adh,ajb}, ranging
from quantum Hall effect \cite{qhe,qhe1,qhe2}, gauge theories
\cite{gt,gt1}, chaos \cite{al,nala,gar}, fractional statistics
\cite{apo,fs1,fs2,fs3,fs4,fs5,fs6} etc. The CSM and SM type
long-range interactions have also manifested in models dealing
with pairing interactions\cite{pi} and phase transitions\cite{pt}.
It is known that, the CSM is related to random matrix theory
\cite{sut1,al,nala,dys,rmt2,rmt3} and enables one to capture the
universal aspects of various physical phenomena \cite{al}. The
Brownian motion model of Dyson connects the random matrix theory
(RMT) with exactly solvable models \cite{dys}. The role of RMT in
the description of the level statistics of chaotic systems is
well-known \cite{bogi,bogi1}.

Some time back, a short-range Dyson model was introduced to
understand the spectral statistics of systems, which are
``non-universal with a universal trend'' \cite{trend}. It is known
that, there are dynamical systems which are neither chaotic nor
integrable, the so called pseudo-integrable systems which exhibit
the above mentioned level statistics \cite{pse}. Aharanov-Bohm
billiards \cite{ab}, three dimensional Anderson model at the
metal-insulator transition point \cite{mit} and some polygonal
billiards \cite{bg} fit into the above description. Quite
recently, a new class of one dimensional, exactly solvable
many-body quantum mechanical models on the line, with nearest and
next-to-nearest neighbour interactions, have been introduced
\cite{ak1,ak2}, which are related to this short-range Dyson model.
Further, using the symmetrized version of this model, it has been
shown that there exists an off-diagonal long-range order in the
system which indicates the presence of different quantum phases
\cite{ga}. Apart from possessing a good thermodynamic limit {\it
i.e.}, ${\rm lim}_{N \rightarrow \infty} \frac{E_0} {N}$ is
finite, these models are more physical in the sense that, unlike
the CSM and SM, where all particles experience pairwise
interaction of identical strength, irrespective of their
distances, here the interactions are only nearest neighbour and
next-to-nearest neighbour. These models are exactly solvable, but
not integrable. Hence, it is interesting to enquire as to how many
features of the integrable CSM type systems are retained in the
present case. For example, the scattering phenomena is quite
interesting in the CSM case, since the outgoing waves can be shown
to be of the incoming type, with momenta $k_i^\prime = k_{N+1-i}
(i = 1,2, \cdots,N)$, where $k_i$'s are the incoming momenta.
Remarkably, the phase shifts are energy independent, a result
ascribable to the scale invariance of the inverse square
interaction. Since, in the present case also, scale invariance
holds and the interaction goes to zero as the particle separation
increases, it is of deep interest to study the scattering
phenomena. Similarly, for the bound state problem, CSM can be
exactly made equivalent to a set decoupled oscillators, via a
similarity transformation \cite{prb}. Hence, it is also of
interest to check the same here to understand the precise
differences in the Hilbert space structure between integrable and
non-integrable Hamiltonians, possessing identical spectra. It
should be mentioned that, in the present case, the degeneracy is
less, since the lack of quantum integrability i.e., a desired set
of mutually commuting operators having common eigenfunctions with
the Hamiltonian, reduces the degeneracy. Also, these models, being
of recent origin, need to be analyzed thoroughly, in order to
unravel their properties, as has been done for the CSM, SM and
their generalizations.

The present paper is devoted to an investigation of these models,
and deals with, both the scattering and the bound state problems.
It is organized as follows: In Sec.II, we study the scattering
problem and show that the scattering state is a coherent state.
The connection between the scattering Hamiltonian and that of the
free particles in this model is then demonstrated. We then study
the scattering phase shift and point out its similarities and
differences with the Calogero case. In Sec.III, we analyze the
relationship of the bound state problem with the decoupled
oscillators and find some wavefunctions explicitly in the
Cartesian basis. This analysis reveals explicitly that the
degeneracy of this model is less as compared to the CSM. Finally,
in Sec.IV, the nearest neighbour and next-to-nearest neighbour
$A_{N - 1}$ and $BC_{N}$ models on the circle are studied; a part
of their excitation spectra is obtained through symmetry
arguments.
\section{The Scattering Problem}

\subsection{Realization of the scattering states as coherent states}
The scattering Hamiltonian, with nearest and next-to-nearest
neighbour, inverse square interactions, in the units $\hbar =  m =
1$, is given by,
\begin{equation} \label{sca}
H_{sca} = - \frac{1}{2} \sum_{i=1}^N \frac{{\partial}^2}{\partial
x_i^2} + \beta (\beta - 1) \sum_{i=1}^{N} \frac {1}{(x_i -
x_{i+1})^2} - \beta^2 \sum_{i=2}^{N} \frac {1}{(x_{i-1} - x_i)
(x_i - x_{i+1})}\qquad,
\end{equation}
here $x_{N + i} = x_i$. The corresponding bound state Hamiltonian,
contains an additional oscillator potential:
\begin{equation}\label{bs} H = - \frac{1}{2}
\sum_{i=1}^N \frac{{\partial}^2}{\partial x_i^2} + \frac{1}{2}
\sum_{1=1}^N x_i^2 + \beta (\beta - 1) \sum_{i=1}^{N} \frac
{1}{(x_i - x_{i+1})^2} - \beta^2 \sum_{i=1}^{N} \frac {1}{(x_{i-1}
- x_i) (x_i - x_{i+1})}\,\,.
\end{equation}
It is interesting to note that, the scattering eigenstates can be
constructed as coherent states of the bound state eigenfunctions
like that of the Calogero case \cite{coherent}. To be precise, we
show that the polynomial part of the bound-state wavefunctions
enter into the construction of the scattering states. For this
purpose, we first identify an $SU(1,1)$ algebra containing the
bound and scattering Hamiltonians as its elements, after suitable
similarity transformations:
\begin{eqnarray} Z^{-1} (- H_{sca}) Z &=& \frac{1}{2} \sum_{i=1}^N
\frac{{\partial}^2}{\partial x_i^2} + \beta \sum_{i=1}^{N} \frac
{1}{(x_i - x_{i+1})}(\partial_i - \partial_{i+1})
\equiv T_+ \quad,\nonumber\\
\hat{S}^{-1} (- H/2) \hat{S}  &=& -\frac{1}{2} \left( \sum_i
x_i \partial_i + E_0 \right) \equiv T_0 \quad.\nonumber\\
\end{eqnarray}
Here, $Z \equiv \prod_{i=1}^N \mid x_i - x_{i+1} \mid^\beta$ and
$\hat{S} \equiv \exp\{- \frac{1}{2} \sum_i x_i^2\}\,\,
Z\,\,\exp\{- \frac{1}{2} T_+\}$\,\,.\\
Defining,
\begin{eqnarray}
\frac{1}{2} \sum_i x_i^2 \equiv T_- \qquad,
\end{eqnarray}
one can easily check that $T_{\pm}$ and $T_0$ satisfy the usual
$SU(1,1)$ algebra:
$$
[ T_+ , T_- ] = - 2 T_0 \quad,\quad [ T_0 , T_{\pm} ] = \pm
T_{\pm} \quad .
$$
The quadratic Casimir for the above algebra is given by,
\begin{eqnarray}\label{qc}
\hat{C} = T_- T_+ - T_0 (T_0 + 1) = T_+ T_- - T_0 (T_0 - 1)\quad.
\end{eqnarray}

By finding a canonical conjugate of $T_+$ \cite{cc,gps,sun}:
\begin{equation} \label{vs}
[ T _+,  \tilde{T_-} ] = 1 \qquad,
\end{equation}
one can construct the coherent state $<x \mid m, k>$, the
eigenstate of $T_+$ \cite{cs-amp}, which are nothing but the
scattering states,
\begin{equation} \label{coher} <x \mid m, k> =
U^{-1} P_m(x) = e^{- \frac{1}{2} k^2 \tilde {T_-}} P_m(x)\quad,
\end{equation}
with
\begin{eqnarray}\label{3}
T_+ P_m(x) \equiv \bigg [ \frac{1}{2} \sum_{i=1}^N
\frac{{\partial}^2}{\partial x_i^2} + \beta \sum_{i=1}^{N} \frac
{1}{(x_i - x_{i+1})}(\partial_i -
\partial_{i+1}) \bigg ] P_m (x) = 0 \, .
\end{eqnarray}
It is known that this equation admits homogeneous solutions
\cite{ak2} {\it i.e}, $T_0 P_m(x) = - [(m + E_0)/2] P_m(x)$. Here
$m$ refers to the degree of homogeneity of $P_m(x)$. It is easy to
see that, $U^{- 1} P_m(x)$ is the eigenstate of $T_+$. Starting
from $T_+ P_m(x) = 0$, one gets,
\begin{eqnarray}
U^{-1} T_+ U U^{-1} P_m(x) = 0\quad,
\end{eqnarray}
{\it i.e},
\begin{equation}
T_+ U^{-1} P_m(x) = - \frac{1}{2} k^2 U^{-1} P_m(x) \quad.
\end{equation}
The scattering state is given by, $\psi_{sca} = Z U^{- 1}
P_m(x)$\,\,, since $H_{sca} = - Z T_+ Z^{- 1}$, therefore,
$H_{sca} \psi_{sca} = \frac{k^2}{2}
 \psi_{sca}$.
To find $<x \mid m, k>$ explicitly, we have to determine
$\tilde{T_-}$. By choosing $\tilde{T_- } = T_-  F(T_0)$, Eq.
(\ref{vs}) becomes
\begin{eqnarray}
[ T _+ ,  T_- F(T_0) ] =
F(T_0) T_+ T_- - F(T_0 + 1 ) T_- T_+ = 1\quad, \nonumber\\
F(T_0) \{ \hat C + T_0 (T_0  - 1 )\} - F ( T_0 + 1 ) \{ \hat C +
T_0 ( T_0 + 1)\} = 1 \qquad,
\end{eqnarray}
yielding,
\begin{equation}\label{ft0}
F(T_0 ) = \frac{-T_0 + a}{\hat C + T_0 (T_0 - 1)} \qquad.
\end{equation}
Here, $a$ is a parameter to be fixed along with the value of the
quadratic Casimir $\hat C$, by demanding that the above commutator
is valid in the eigenspace of $T_0$. Eq. (\ref{vs}) when used on
$P_{m}(x)$, yields, $a = 1 - (E_0 + m)/2$. Similarly,
\begin{eqnarray}
\hat{C} P_m(x) = (T_- T_+ - T_0 (T_0 + 1)) P_m(x) = C P_m
(x)\quad,\nonumber\\
\end{eqnarray}
where, $C = \frac{1}{2}(m + E_0) (1 - (m + E_0)/2)$. One then
finds,
\begin{eqnarray}
F(T_0) P_m(x) &=& \frac{ - T_0 + a}{C + T_0 (T_0 - 1)} P_m(x) \nonumber\\
      &=& - \frac{1}{ T_0 - (m + E_0)/2} P_m(x) \qquad.
\end{eqnarray}
Explicitly, we have,
\begin{eqnarray} \label{suppu}
< x \mid m, k > &=& e^{- \frac{1}{2} k^2 \tilde{T_-}} e^{-T_+}
P_m(x)
 \nonumber\\
&=& e^{ - \frac{1}{4} k^2 } e^{-T_+} e^{- \frac{1}{2} k^2 \tilde
{T_-}}
P_m(x) \nonumber\\
&=& e^{ - \frac{1}{4} k^2} \sum_{n=0}^\infty \frac{(k^2/2)^n}{(E_0
+ m
+ n)!} L_n^{(E_0 - 1 + m)}(r^2/2) P_m(x) \nonumber\\
&=& e^{- k^2/4} (k/2)^{- (E_0 - 1 + m)}  (r)^{- (E_0 - 1 + m)}
J_{E_0 - 1 + m}(k r) P_m(x) \quad,
\end{eqnarray}
where, $r^2 = \sum_i x_i^2$ and $J_{E_0 - 1 + m}(k r)$ is the
Bessel function. Note that, there is an additional factor of $e^{-
T_+}$ in the above equation, this has been introduced for
calculational convenience and does not alter our results, since
$e^{- T_+} P_m = P_m$. In order to arrive at the above result, we
have made use of the following:
$$T_+ \left(r^{2 n} P_m(x) \right) = 2 n (E_0 - 1 + m + n) r^{2
(n - 1)} P_m(x) \qquad,$$ and also the identity \cite{gra},
$$J_\alpha(2 \sqrt{x z}) e^z (xz)^{- \alpha/2} = \sum_{n=0}^\infty
\frac{z^n}{(n + \alpha + 1)!} L_n^\alpha(x)\quad.$$

Note that, the above wavefunction can also be obtained by solving
the scattering Hamiltonian explicitly. However, we have chosen
this algebraic method, since it will be of subsequent use.

\subsection{Connection to Free Particles}

The fact that as in the Calogero case, the spectrum of the
scattering Hamiltonian matches with that of the free particles and
that, the phase shift, as will be shown later, are energy
independent, suggests a possible connection of this system with
free particles\cite{gon}. We now show the same by making use of
the algebraic structures already introduced.

The following generators,
\begin{eqnarray}
K_+ = H_{sca},\qquad \tilde{K_+} = H_{sca}(\beta =
0)\,\,,\nonumber
\end{eqnarray}
\begin{eqnarray}
K_- = \frac{1}{2} \sum_{i = 1}^N x_i^2 = \tilde{K}_-\,\,,\nonumber
\end{eqnarray}
and
\begin{eqnarray}
K_0 = - \frac{i}{4} \sum_{i = 1}^N (2 x_i \partial_i + 1) =
\tilde{K}_0,
\end{eqnarray}
satisfy $[K_0,\,\,K_{\pm}(\tilde{K}_{\pm})] = \pm i K_{\pm}
(\tilde{K}_{\pm})$,\,\,$[K_- (\tilde{K}_-)$,\,\,$ K_+
(\tilde{K}_+)] = 2 i K_0 (\tilde{K}_0)$. The normalization of the
generators have been chosen differently for convenience. It can be
verified that, the following operator,
\begin{eqnarray}
U = e^{\frac{i\pi}{2} (\tilde{K}_+ + \tilde{K}_-)} e^{-
\frac{i\pi}{2} (K_+ + K_-)},\nonumber
\end{eqnarray}
maps the $H_{sca}$, with interactions, to a interaction-free
system, $\it i.e.$,
\begin{eqnarray}
U\,\,H_{sca}\,\,U^\dagger = H_{sca} (\beta = 0) = - \frac{1}{2}
\sum_i \frac{\partial^2}{\partial x_i^2}\qquad.
\end{eqnarray}
This result motivates one to analyze, explicitly, the precise
correspondence of the respective Hilbert spaces of the interacting
and non-interacting systems. Care has to be taken to ensure that,
the mapped states are members of the Hilbert space. This analysis,
for the present case, as well as for the Calogero model has not
been carried out so far. We hope to address this problem in the
near future.

\subsection{Analysis of the Scattering Phase Shift}

As is well known, in order to obtain the scattering phase shifts,
one has to analyze the asymptotic behaviour of the eigenfunctions
of the Hamiltonian, when all particles are far apart from each
other. In the present case, the most general stationary
eigenfunction can be written as a superposition of all the $m$
dependent states as,
\begin{eqnarray} \label{sta}
\psi = Z \sum_{m = 0}^\infty \sum_q C_{m,q} \,\, (r/2)^{- (E_0 - 1
+ m)} J_{E_0 - 1 + m}(k r) P_{m,q}(x) \qquad,
\end{eqnarray}
where, $C_{m,q}$'s are some coefficients and the index $q$ refers
to the number of independent $P_{m}$'s at the level $m$.

The asymptotic limit of Eq. (\ref{sta}) can be obtained from the
asymptotic behaviour of the Bessel function $J_{E_0 - 1 + m}(k r)$
and from the fact that $r^{- m} P_{m, q} (x)$ does not change in
this limit, since $P_{m, q}$'s are homogeneous functions of degree
m. One obtains,
\begin{eqnarray}
\psi \sim \psi_{\rm{in}} + \psi_{\rm{out}} \qquad,
\end{eqnarray}
where,
\begin{eqnarray} \label{ain}
\psi_{\rm{in}} \equiv e^{i (E_0 - 1) \pi/2} (2 \pi k r)^{-
\frac{1}{2}} \left( Z  r^{- (E_0 - 1 )}\sum_{m = 0}^\infty \sum_q
C_{m,q} \,\, (r/2)^{- m} e^{i m \pi/2} P_{m,q}(x)\right) e^{- i k
r}\,\,,
\end{eqnarray}
and
\begin{eqnarray} \label{aout}
\psi_{\rm{out}} \equiv e^{- i (E_0 - 1) \pi/2} (2 \pi k r)^{-
\frac{1}{2}} \left( Z  r^{- (E_0 - 1 )} \sum_{m = 0}^\infty \sum_q
C_{m,q} \,\, (r/2)^{- m} e^{- i m \pi/2} P_{m,q}(x) \right) e^{i k
r}\,\,.
\end{eqnarray}
The crucial differences between the Calogero and the present model
arise from this point. Denoting the equivalent of $P_{m,q} (x)$'s
in the Calogero case, as $\tilde{P}_{m,q}(x)$'s, we first describe
the Calogero scattering problem, and then compare the results of
this model with the same. Though the $\tilde{P}_{m,q} (x)$'s,
which are the solutions of the generalized Laplace equation
\cite{cal}, have not been found explicitly so far, for arbitrary
$m$, Calogero had shown the existence and completeness of these
homogeneous symmetric polynomials. Hence, by choosing suitable
values for the coefficients $C_{m,q}$'s, one could characterize
the incoming wave, for the Calogero case as,
$$
\psi_{\rm{in}} \equiv c \exp\left\{ i \sum_i k_i x_i  \right\}
\quad,
$$
with $k_i \le k_{i+1}, i = 1, 2, \cdots, N - 1$, as the stationary
eigenfunction in the center-of-mass frame, {\it i.e.}, $\sum_i k_i
= 0$. Further, since, the $\tilde{P}_{m,q}$'s are symmetric under
cyclic permutations and are homogeneous, one obtains,
$\tilde{P}_{m,q} (- T x) = e^{- i m \pi} \tilde{P}_{m,q}(x)$.
where, $T$ denotes a cyclic permutation of the particle
coordinates. Hence, $\psi_{\rm{out}}$ for the Calogero model can
be written as,
\begin{eqnarray}\label{aout1}
\psi_{\rm{out}} &\equiv& e^{- i (E_0 - 1) \pi} (2 \pi \bar{k}
r)^{- \frac{1}{2}} \left( Z  r^{- (E_0 - 1 )} \sum_{m = 0}^\infty
\sum_q C_{m,q} \,\, (r/2)^{- m} e^{i m \pi/2} \tilde{P }_{m,q}(- T
x) \right) e^{- i \bar{k} r} \,\,,
\end{eqnarray}
where, $\bar{k} = - k$, and $- T x_i = - x_{N+1-i}$. The action of
$(- T)$ takes a given particle ordering $x_i \ge x_{i+1}$ to $ -
x_{N+1-i} \ge - x_{N-i}$, and hence preserves the order {\it
i.e.}, the sector of the configuration space. This is the reason,
invariance of $\tilde{P}_{m,q}(x)$'s under cyclic permutation is
enough to compute the phase shifts. Comparison with
$\psi_{\rm{in}}$, yields,
\begin{eqnarray}\label{10}
\psi_{\rm{out}} &=& c e^{- i (E_0 - 1) \pi} \exp\left\{ - i \sum_i
\bar{k}_i T x_i  \right\} \nonumber\\
&=& c e^{- i (E_0 - 1) \pi} \exp\left\{ i \sum_i k_{N +1 - i} x_i
\right\} \quad,
\end{eqnarray}
where, we have used $\bar{k}_i = - k_i$ and the cyclic permutation
has been carried on the momentum variables. From above, it is
clear that, the initial scattering situation characterized by the
initial momenta, $k_i (i = 1, 2, \cdots, N)$, goes over to the
final configuration, characterized by the final momenta,
$k_i^\prime = k_{N+1-i}$ and the phase shifts are energy
independent. In the present model, $P_{m,q} (x)$'s are the
solutions of Eq. (\ref{3}), which is symmetric only under cyclic
permutations. Hence, one needs to check if the same steps as
narrated above for Calogero's case also applies here. First of
all, we should find out if the number of $P_{m,q}$'s are same
here. For the sake of clarity, we first consider the four particle
case and concentrate on the homogeneous solutions of degree four.
Since the monomial symmetric functions, provide a linearly
independent basis set, we can expand $P_4 (x_i)$ as,
\begin{eqnarray}
P_4 = a \sum_{i = 1}^4 x_i^4 + b \sum_{i \ne j}^4 x_i^3 x_j + c
\sum_{i < j}^4 x_i^2 x_j^2 + d \sum_{i \ne j \ne l}^4 x_i^2 x_j
x_l + e \sum_{i < j < l < p}^4 x_i x_j x_l x_p\qquad.
\end{eqnarray}
The operation of $T_{+}$ on $P_4$ gives the following three sets
of conditions,
\begin{eqnarray}
&&[2 (3 + 4 \beta) a - 2 \beta b + (3 + 4 \beta) c - 2 \beta
d]\,\, \sum_{i = 1}^4 x_i^2 = 0\,\,,\nonumber\\
&&[4 \beta a + 2 (3 + 4 \beta) b - 2 \beta c + 2 (1 - \beta) d -
\beta e]\,\,
\sum_{i = 1}^4 x_i x_{i + 1} = 0\,\,,\nonumber\\
&&[6 (1 + 2 \beta) b + 2 (1 - 2 \beta) d]\,\,\sum_{i = 1}^4 x_i
x_{i + 2} = 0\qquad,
\end{eqnarray}
where a, b, c, d and e are unknown constants to be determined from
the above equations. For the purpose of comparison, in the
Calogero case,
\begin{eqnarray}
\tilde {T}_+ \tilde{P}_4 (x) \equiv \left[\frac{1}{2}\sum_{i =
1}^4 \frac{\partial^2}{\partial x_i^2} + \beta \sum_{i<j}^4
\frac{1}{(x_i - x_j)} (\partial_i - \partial_j)\right]
\tilde{P}_4(x) = 0\qquad,
\end{eqnarray}
yields,
\begin{eqnarray}
&&[6 (1 + 2 \beta) a - 3 \beta b + 3 (1 + 2 \beta) c - 3 \beta
d]\,\, \sum_{i = 1}^4 x _i^2 = 0\,\,,\nonumber\\
&&[4 \beta a + 2(3 + 2 \beta) b - 2 \beta c + 2 (1 - \beta) d -
\beta e]\,\, \sum_{i < j}^4 x_i x_j = 0\qquad.
\end{eqnarray}

Hence, the number of solutions are less in the present model as
compared to the Calogero case. This reduction in the number of
solutions takes place because, the interaction term contained in
the $T_{+}$ operator, leads to the split of the monomial symmetric
functions of degree $(m - 2)$, giving additional conditions unlike
the Calogero case. Explicit calculations for a number of few body
examples yields similar results\cite{ak2}. Since the number of
$P_{m,q}$'s also represent degeneracy, the same is less here.
Hence, it is clear that, $P_{m,q}$'s do not form a complete set.
All the solutions obtained so far are symmetric and belongs to a
subset of the Calogero case. The reason of the symmetric nature of
the polynomials lies in the interaction term in $T_+$. In order
that the action of the interaction term, $\beta \sum_{i = 1}^N
\frac{1}{x_i - x_{i + 1}} (\partial_i -
\partial_{i + 1})$, on the polynomial $P_{m}(x)$ results in a
polynomial of degree $m - 2$, the denominator needs to be
cancelled. This results in the symmetrization of the polynomial,
since the nearest neighbour couplings arising from $\frac{1}{x_i -
x_{i + 1}} $ connects each particle with every other member of the
set. The reduction in the number of symmetric polynomials can also
be understood from the point of view of integrability. Since
Calogero model is fully integrable, there are desired number of
operators, commuting with the Hamiltonian, which can be used for
connecting the members of a given set of $P_{m,q}$'s akin to the
angular momentum raising and lowering operators in the central
force problem. The fact that, the number of $P_{m,q}$'s are less
here indicates that, the corresponding commuting constants of
motion are less here. This point will be further elaborated in the
next section.

As has been mentioned earlier, the completeness of the solutions
of the generalized Laplace equation in the Calogero model enables
one to relate all the partial waves of the incoming state with
those of the outgoing state, with constant energy independent
phase shifts. As is clear from Eqs. (\ref{ain}), (\ref{aout}) and
(\ref{aout1}), in the present case, the outgoing wavefront can be
made to look like the incoming wavefront. Hovever, only those
partial waves, which are solutions of Eq. (\ref{3}) will acquire
energy independent phase shifts $e^{- i (E_0 - 1)\pi}$, the rest
will be unaffected by the interaction. This difference between the
Calogero and the present model arises, because of the reduction of
the number of $P_{m,q}$'s here.

\section{THE BOUND STATE PROBLEM}

\subsection{Mapping of the model to decoupled oscillators}
The bound state Hamiltonian is given by,
\begin{equation}\label{bs}
H = - \frac{1}{2} \sum_{i=1}^N {\partial}_i^2 + \frac{1}{2}
\sum_{1=1}^N x_i^2 + \beta (\beta - 1) \sum_{i=1}^{N} \frac
{1}{(x_i - x_{i+1})^2} - \beta^2 \sum_{i=1}^{N} \frac {1}{(x_{i-1}
- x_i) (x_i - x_{i+1})}\,\,,
\end{equation}
where, $\partial_i \equiv \frac{{\partial}}{\partial x_i}$ and
$x_{N+i} \equiv x_i$. It is worth pointing out that, for three
particles, this model is equivalent to the CSM, since the three
body term vanishes in this case. The ground-state wavefunction and
the energy of this system \cite{ak1,ak2} are respectively given
by, $\psi_0 = G\, Z\,$ and $E_0 = (N/2 + N \beta)$, where $G$ and
$Z$ have been defined earlier. The first term of $E_o$ is the
ground state energy of $N$ oscillators and the second one comes
from the interaction.

In the following, we make use of a method developed in
Ref.(\onlinecite{prb}) to show the equivalence of this model to a
set of decoupled oscillators. For that purpose, we perform a
similarity transformation on the Hamiltonian, by its ground-state
wavefunction, to yield
\begin{eqnarray}
{H}^\prime \equiv \psi_0^{-1} H \psi_0 = \sum_i x_i \partial_i +
E_0 - T_+ \qquad,
\end{eqnarray}
where, $ T_+ \equiv \frac{1}{2} \sum_{i=1}^N
\frac{{\partial}^2}{\partial x_i^2} + \beta \sum_{i=1}^{N} \frac
{1}{(x_i - x_{i+1})}(\partial_i -
\partial_{i+1})$.
Here, we confine ourselves to a sector of the configuration space
given by $x_1 \ge x_2 \ge \cdots \ge x_{N-1} \ge x_N$. Using the
identity,
$$[\sum_i x_i \partial_i \,\,,\,\,e^{- T_+/2}] = T_+ e^{- T_+/2} \qquad,$$
it is easy to see that
\begin{eqnarray} \label{euler}
\bar{H} \equiv e^{T_+/2} {H}^\prime e^{- T_+/2} = \sum_i x_i
\partial_i + E_0 \qquad.
\end{eqnarray}
>From the above diagonalized form, it is evident that, the spectrum
of $H$ is like that of $N$ uncoupled oscillators and is linear in
the coupling parameter $\beta$. Explicitly, $\bar{H}$ can be made
equivalent to the decoupled oscillators:
\begin{eqnarray}
G \,\,e^{- T_+(\beta=0)/2}\,\, \bar{H}\,\, e^{T_+(\beta=0)/2}\,\,
G^{-1} = - \frac{1}{2} \sum_{i=1}^N {\partial}_i^2 + \frac{1}{2}
\sum_{1=1}^N x_i^2 + E_0 - N/2 \qquad.
\end{eqnarray}
Making inverse similarity transformations, one can write down the
raising and lowering operators for $H$, akin to the CSM. However,
the eigenfunctions of $H$ can be constructed straightforwardly by
making use of Eq. (\ref{euler}); since the eigenfunctions of
$\sum_i x_i \partial_i$ are homogeneous polynomails of degree $n$
in the particle coordinates, $n$ being any integer. Although, the
similarity transformation formally maps the Hilbert space of the
interacting problem to that of the free oscillators, it needs a
careful study. The singular terms present in $T_+$ may yield
states, which are not members of the Hilbert space. Below, we
clarify this point.

\subsection{Eigenfunctions in the Cartesian Basis}

In the following, we present some eigenfunctions for the
$N$-particle case, computed using the power-sum basis, $P_l(x) =
\sum_{i=1}^N x_i^l$, {\i.e.}, ${\psi_l = \psi_0 S_l}$; here, $S_l
\equiv e^{- T_+/2} P_l$, and the corresponding energy eigenvalue
is $E_l = (l + E_0)$, $l$ being an integer.

The wavefunction (unnormalized) corresponding to the
center-of-mass degree
 of freedom, $R
= \frac{1}{N} \sum_{i=1}^N x_i$, is found to be (we use the
notation $\psi_{n_1,n_2,\cdots, n_N} = \psi_0\,\,\exp\{-
\frac{1}{2} {T_+}\} \prod_l P_l^{n_l}$),
\begin{equation}
\psi_{n_1,0,0,\cdots} = \psi_0\,\,\exp\{- \frac{1}{2} {T_+}\}
R^{n_1} = \psi_0 \,\,\exp\{- \frac{1}{4} \sum_{i=1}^N
\partial_i^2\} \,\,R^{n_1}\qquad.
\end{equation}
This can be cast in the form \cite{ng3},
\begin{equation}
\psi_{n_1,0,0,\cdots} = c \,\, \psi_0\,\,\sum_{\sum_{i=1}^N m_i =
n_1} \prod_{i=1}^N \frac{H_{m_i}(x_i)}{m_i!} \qquad,
\end{equation}
where, $H_{m_i}(x_i)$'s are the Hermite polynomials, and c is a
constant. Similarly, the eigenfunction for the radial degree of
freedom, $r^2 = \sum_i x_i^2$, can be obtained from,
\begin{equation}
\psi_{0,n_2,0, \cdots} = \psi_0\,\,\exp\{- \frac{1}{2} {T_+}\}
\,\,(r^2)^{n_2} = \psi_0 e^{- \frac{1}{2} T_+}P_2^{n_2} \qquad.
\end{equation}
For the sake of clarity, we give below the explicit derivation for
$\psi_{0, 1, 0, 0}$ for the four particle case and then generalize
the result for arbitrary number of particles and levels. For four
particles,
\begin{eqnarray}
\psi_{0, 1, 0, 0} = \psi_0\,\, e^{- \frac{1}{2} T_+} P_2 =
\psi_0\,\, e^{- \frac{1}{2} T_+} (x_1^2 + x_2^2 + x_3^2 +
x_4^2)\quad.
\end{eqnarray}
We note that, $T_+ P_2 = 4 (2 \beta + 1)$\,\,, $T_+^2 P_2 = 0$,
and hence, $\psi_{0, 1, 0, 0} = \psi_0 (P_2 - 2 (2 \beta + 1))$.
For $N$ particles, it can be verified that,\,\, $T_+ r^{2 n} = 2 n
(E_0 - 1 + n) r^{2 (n - 1)}$ and this gives $\psi_{0,{n_2},0,
\cdots}$ as \cite{ng4},
\begin{eqnarray}
\psi_{0,n_2,0, \cdots} &=& \psi_0\,\, \sum_{m=0}^{n_2} \frac{(-
1)^m}{m! (n_2 - m)!} \frac{(E_0 - 1 + n_2)!}{(E_0 - 1 + m)!}
{(r^2)}^{m}
\qquad,\nonumber\\
&=& c \,\, \psi_0\,\,L_{n_2}^{E_0 - 1}(r^2) \qquad,
\end{eqnarray}
where, $L_{n_2}^{E_0 - 1}(r^2)$ is the Lagurre polynomial.

Further, for four particles, $S_3 = P_3 - \frac{3}{2} (2 \beta +
1) P_1$. However, it can be checked that, $S_4 = e^{- T_+/2} P_4$,
does not terminate as a polynomial and results in a function with
negative powers of the particle coordinates, which is not
normalizable with respect to the ground-state wavefunction as a
measure. This indicates that, there is less degeneracy in the
present model, as compared to the symmetrized states of the
decoupled oscillators. Finding the other wavefunctions explicitly,
for an arbitrary number of particles, and also the exact
degeneracy structure of this model remains an open problem. In the
above, we have concentrated in finding the wavefunctions in the
Cartesian basis; the interested readers are refererred to
Ref.(\onlinecite{ak1}) for some wavefunctions in the angular
basis.

Below, we list a few eigenfunctions constructed by using the
elementary symmetric functions (we follow the notations of
Ref.(\onlinecite{jack}) for the symmetric polynomials).
\begin{eqnarray}\label{4}
e_1 = \sum_{1 \le i \le N} x_i \,,\,\, e_2 = \sum_{1 \le i < j \le
N} x_i x_j \,,\,\, e_3 = \sum_{1 \le i < j < k \le N} x_i x_j x_k,
\cdots, e_N = \prod_{i=1}^N x_i \, .
\end{eqnarray}
In this case, $\psi_{\{m_i\}} = \psi_0 B_{\{m_i\}},\quad$
$B_{\{m_i\}} = e^{- T_+/2} \prod_i (e_i)^{m_i}$, \,\, and the
corresponding eigenvalues are, $E_{\{m_i\}} = \sum_i i m_i + E_0$.
Some of the $B_{\{m_i\}}$'s for the four particle case are listed
below:
\begin{eqnarray}
B_{2,0,0,0} &=& e_1^2 - 2, \,\,B_{1,1,0,0} = e_1 e_2 - \frac{1}{2}
(3 - 4
\beta) e_1, \,\,B_{3,0,0,0} = e_1^3 - 6 e_1, \nonumber \\
B_{2,1,0,0} &=& e_1^2 e_2 - (3 - 2 \beta) e_1^2 - 2 e_2 +  (3 - 4
\beta),
B_{4,0,0,0} = e_1^4 - 12 e_1^2 + 12, \nonumber\\
B_{3,1,0,0} &=& e_1^3 e_2 - \frac{1}{2}(9 - 4 \beta) e_1^3 - 6 e_1
e_2 + 6
(3 - 2 \beta) e_1, \,\,B_{5,0,0,0} = e_1^5 - 20 e_1^3 + 60 e_1, \nonumber 
\\
B_{4,1,0,0} &=& e_1^4 e_2 - 2 (3 -  \beta) e_1^4 - 12 e_1^2 e_2 +
12 e_2 + 6 (9 -4 \beta ) e_1^2 -12 (3 -2 \beta)\,\,.
\end{eqnarray}

At this point, it is worth recollecting the Stanley-Macdonald
conjecture \cite{jack}, which states that, the coefficients of the
interaction parameter $\beta$ are positive integers, when the Jack
polynomials \cite{jack} are expressed in terms of the monomial
symmetric functions with a suitable normalization. This conjecture
was later proved by Sahi \cite{sahi}. Similar feature appears in
the case of the Hi-Jack polynomials \cite{roj}, which are the
polynomial part of the wavefunctions of the CSM, but with an
exception that the coefficient $\beta$ can also be negative.
Remarkably, from the above explicit computations of the
polynomials, we also find that the coefficients of the interacting
parameter $\beta$ are integers (both positive and negative),
though we have used elementary symmetric functions. It will be
interesting to check whether the modified Stanley-Macdonald
conjecture also holds in the present case for $N$ particles.

\section{Model on a Circle}

\subsection{$A_{N-1}$ Model}

Recently, Jain et al. \cite{ak1,ak2} have studied a model with
nearest and next to nearest neighbour interactions and with
periodic boundary conditions as given by
\begin{eqnarray}\label{cir}
H = &-& \frac{1}{2} \sum_{j} {\partial}_j^2 + \beta (\beta - 1)
\frac{\pi^2}{L^2} \sum_{j}\frac{1}{\sin^2[\frac{\pi}{L}(x_j
- x_{j + 1})]} \nonumber\\
&-& \beta^2 \frac{\pi^2}{L^2}\sum_{j} \cot[\frac{\pi}{L}(x_{j - 1}
- x_j)] \cot [\frac{\pi}{L}(x_j - x_{j + 1})] \, ,
\end{eqnarray}
with $x_{i+N} = x_i$. They have shown that the ground state energy
eigenvalue and the eigenfunction for this model are given by
\begin{eqnarray}\label{5.2}
\psi_0 = \prod_{j}^N [\sin \frac{\pi}{L} (x_{j} -
 x_{j + 1})]^\beta \, , E_0 = N\beta^2 \frac{\pi^2}{L^2} \, .
\end{eqnarray}
The purpose of this section is to obtain a part of the excitation
spectrum of this model. To that end, we substitute
\begin{eqnarray}\label{5.3}
\psi = \psi_0 \phi \, ,
\end{eqnarray}
in the eigenvalue equation for the Hamiltonian. It is then easily
shown that $\phi$ satisfies the equation
\begin{eqnarray}\label{5.4}
\left(- \frac{1}{2} \sum_{j = 1}^N \partial_j^2 -  \beta
\frac{\pi}{L} \sum_{j = 1}^N
 \left[\cot \frac{\pi}{L}
(x_j - x_{j + 1}) - \cot \frac{\pi}{L}(x_{j - 1} - x_j)\right]
\partial_j + E_0 - E \right) \phi = 0 \, .
\end{eqnarray}
Introducing,  $z_j = \exp(2 i \pi x_j/L)$, Eq. (\ref{5.4}) reduces
to
\begin{eqnarray}\label{5.4a}
H_1 \phi = (\epsilon - \epsilon_0) \phi
\end{eqnarray}
where
\begin{eqnarray}\label{5.5}
H_1 = \sum_{j= 1}^N D_j^2 + \beta \sum_{j = 1}^N \left [\frac{z_j
+z_{j+1}}{z_j - z_{j+1}}\right] (D_j - D_{j+1}) \, ,
\end{eqnarray}
with $D_j \equiv z_j \frac{\partial}{\partial z_j}$ and $\epsilon
- \epsilon_0 = (E- E_0)\frac{L^2}{2\pi^2}$. It is worth pointing
out that the Eqs. (\ref{5.4a}) and (\ref{5.5}) are structurally
similar to those in the SM except $z_k$ is replaced by $z_{j+1}$
in our case.

It may be noted that $H_1$ commutes with the momentum operator $P
= \frac{2\pi}{L} \sum_{i=1}^N z_i \frac{\partial}{\partial z_i}$.
Hence $\phi$ is also an eigenstate of the momentum operator, i.e.,
\begin{eqnarray}\label{5.7}
P \phi = \kappa \phi \, .
\end{eqnarray}
Further, if $\phi$ is an eigenstate of $H_1$ and $P$ then
\begin{eqnarray}\label{5.8}
\phi' = G^{q} \phi \, ,  \ \ G = \Pi_{i=1}^{N} z_i \, ,
\end{eqnarray}
is also an eigenstate of $H'$ and P with eigenvalues $\epsilon -
\epsilon_0 + N q^2 +2 q\kappa$ and $\kappa + N q$ respectively.
Here $q$ is any integer (both positive and negative). Note that
the multiplication by $G$ implements Galilei boost.

It may be noted that the Hamiltonian and hence the $\phi$ equation
is invariant under $z_j \rightarrow z_j^{-1}$. Since, $z_j = e^{2i
\pi x_j/L}$, hence $z_j^{-1} = e^{-2i\pi x_j /L}$ thereby
indicating the presence of left and right moving modes with
momentum $\kappa$ and -$\kappa$. Hence it follows that, if one
obtains a solution with momentum $\kappa$, then by changing $z_j
\rightarrow z_j^{- 1}$, one can get another solution with the same
energy but with the opposite momentum (-$\kappa$). Thus all the
excited states with nonzero momentum are (at least) doubly
degenerate.

Finally, let us discuss the solutions to the $\phi$ equation. So
far we have been able to obtain the following four solutions.
\begin{eqnarray}
(i) \,\, \phi &=& e_1 \, ,  \ \epsilon - \epsilon_0 = 1 +
2\beta,\,\,
\nonumber\\
(ii) \,\, \phi &=& e_{N - 1} \, , \ \epsilon - \epsilon_1 = N - 1 + 
2\beta,\nonumber\\
(iii) \,\, \phi &=& e_1 e_{N-1} - \frac{N}{1+2\beta} e_{N},\,\,
\epsilon - \epsilon_0 = N + 2 + 4\beta,\,\,\nonumber\\
(iv) \,\, \phi &=& e_N \, , \ \epsilon - \epsilon_0 = N\, .
\end{eqnarray}

Here $e_j$ (j=1,2,...,N) denotes the elementary symmetric
functions as defined by Eq. (\ref{4}) (defined in terms of $z_j$).
For example, $e_2 = z_1 z_2 +...+ z_{N - 1}z_N$ and it has N( N -
1)/2 number of terms.

As mentioned above, each of these solution is doubly degenerate.
For example solutions $e_1$ and $e_{N - 1} /e_N$ are degenerate.
By taking the linear combinations of these two complex solutions,
it is easily seen that the two degenerate real solutions are
\begin{eqnarray}
\phi = \sum_{i=1}^N \cos u_i \, , \ \phi = \sum_{i=1}^N \sin u_i
\, ,
\end{eqnarray}
where $u_i = 2\pi x_i /L$. Similarly, all other doubly excited
state solutions can be rewritten as two independent real
solutions. It would appear from this discussion that all the
excited states are doubly degenerate. However, this is not so. In
particular, consider
\begin{eqnarray}\label{5.9}
\phi = \frac{e_1 e_{N-1}}{e_N} - \frac{N}{1+2\beta} \, .
\end{eqnarray}
It is easily shown that it is an exact solution to Eq.
(\ref{5.4a}) with $\epsilon -\epsilon_0 = 2 + 4\beta$ but with
momentum eigenvalue $\kappa =0$. This is a nondegenerate solution
as it remains invariant under $z_i \rightarrow z_i^{-1}$. In terms
of the trignometric functions it can be rewritten as
\begin{eqnarray}
\phi = \sum_{i<j}^N \cos(u_i - u_j) + \frac{N \beta}{1 + 2\beta}
\, .
\end{eqnarray}

At first sight it appears somewhat surprising that whereas in the
Sutherland model, there are so many excited state solutions, in
our case one is able to obtain so few solutions. In this context
it may be noted that whereas the Hamiltonian in the Sutherland
case is invariant under the full permutation group $S_N$, in our
case for $N >3$ the Hamiltonian $H_1$ as given by Eq. (\ref{5.5})
has only cyclic symmetry. If one looks at the solutions to Eq.
(\ref {5.5}), then one finds that the condition of no pole in the
$\beta$-dependent term almost forces $\phi$ to `be invariant under
$S_N$. Now out of the various $e_i$ (i=1,2,...,N), the only ones
in which the demand of cyclic invariance necessarily ensures
invariance under full permutation group  are precisely $e_1,
e_{N-1}, e_N$, in terms of which we have obtained the four
solutions.

\subsection{$BC_{N}$ Model}

Recently Auberson et al. \cite{ak2} have studied a $BC_{N}$ model
with nearest and next-to-nearest neighbour interactions and with
periodic boundary conditions as given by
\begin{eqnarray}\label{4.1}
H &=& - \frac{1}{2} \sum_{i=1}^N \frac{\partial^2}{{\partial
x_i}^2} + \beta(\beta-1)\frac{\pi^2}{L^2} \sum_{i=1}^N
\left[\frac{1}{\sin^2} \frac{\pi}{L}(x_i - x_{i + 1}) +
\frac{1}{\sin^2} \frac{\pi}{L}(x_i + x_{i + 1})\right]\nonumber \\
&-& \beta^2 \frac{\pi^2}{L^2} \sum_{i=1}^N \left[\cot\frac{\pi}{L}
(x_{i - 1} - x_i) - \cot\frac{\pi}{L}(x_{i - 1} + x_i)
\right]\nonumber \\
&&\left[\cot\frac{\pi}{L}(x_i - x_{i + 1}) + \cot\frac{\pi}{L}
(x_i + x_{i + 1})\right] + g_1\frac{\pi^2}{L^2}
\sum_i\frac{1}{\sin^2}\frac{\pi}{L} x_i + g_2\frac{\pi^2}{L^2}
\sum_i \frac{1}{\sin^2}\frac{2\pi}{L} x_i \, .
\end{eqnarray}
Following them, we restrict the coordinates $x_i$ to the sector $L
\ge x_1 \ge x_2 \ge ....\ge x_N \ge 0$. As shown by Auberson et
al. \cite{ak2}, the ground state eigenfunction is given by
\begin{eqnarray}\label{4.2}
\psi_0 = \prod^{N}_{i=1} \sin^{\gamma} \theta_i \prod^N_{i=1}
(\sin^2 2\theta_i)^{\gamma_1 /2} \prod^N_{i=1} [\sin^2 (\theta_i -
\theta_{i + 1})]^{\beta /2} \prod^N_{i=1} [\sin^2 (\theta_i +
\theta_{i + 1})]^{\beta /2} \, ,
\end{eqnarray}
where $g_1,g_2$ are related to $\gamma,\gamma_1$ by
\begin{eqnarray}
g_1 = {\gamma \over 2}[\gamma + 2\gamma_1 - 1] \, , \ \ g_2 =
2\gamma_1 (\gamma_1 - 1) \, .
\end{eqnarray}
The corresponding ground state energy turns out to be
\begin{eqnarray}\label{4.5}
E_0 = {N \pi^2\over 2L^2} (\gamma + 2\gamma_1 + 2\beta)^2 \, .
\end{eqnarray}
By setting one or both of the coupling constants $\gamma,
\gamma_1$ to zero we get the other root systems i.e.
\begin{eqnarray}\label{4.3a}
B_N: \gamma_1 =0 \, , \ \ C_N: \gamma = 0 \, , \ \ D_N: \gamma =
\gamma_1 = 0 \, .
\end{eqnarray}

The purpose of this subsection is to obtain a part of the
excitation spectrum of the $BC_N, B_N, C_N, D_N$ models. To that
end, we substitute
\begin{eqnarray}\label{4.3}
\psi = \psi_0 \phi \, ,
\end{eqnarray}
in the eigenvalue equation for the above Hamiltonian where
$\psi_0$ is as given by Eq. (\ref{4.2}). It is easy to show that
in that case $\phi$ satisfies the equation
\begin{eqnarray}\label{4.4}
&& \bigg [ \sum_{j = 1}^N \frac{\partial^2}{\partial \theta_j^2} +
2 \beta \sum_{j = 1}^N \cot (\theta_j - \theta_{j + 1})
(\frac{\partial}{\partial \theta_j}
-\frac{\partial}{\partial \theta_{j+1}}) \nonumber\\
&& + 2 \beta \sum_{j=1}^{N} \cot (\theta_j + \theta_{j + 1})
(\frac{\partial}{\partial \theta_j} +\frac{\partial}{\partial
\theta_{j+1}}) + 2\gamma \sum_{j=1}^{N} \cot \theta_j
\frac{\partial}{\partial \theta_j}
\nonumber\\
&& + 4\gamma_1 \sum_{j=1}^{N} \cot 2\theta_j
\frac{\partial}{\partial \theta_j} + (E - E_0) \frac{2L^2}{\pi^2}
\bigg ] \phi = 0 \, ,
\end{eqnarray}
where $\theta_j = \frac{\pi x_j}{L}$. Introducing,  $z_j = \exp(2
i \theta_j)$, Eq. (\ref{4.4}) reduces to
\begin{eqnarray}\label{4.4a}
H_1 \phi = (\epsilon - \epsilon_0) \phi
\end{eqnarray}
where
\begin{eqnarray}\label{4.5a}
&& H_1 = \sum_{j= 1}^N D_j^2 +\gamma \sum_{j=1}^{N} \frac{z_j
+1}{z_j -1} D_j
+2\gamma_1 \sum_{j=1}^{N} \frac{z_j^2 +1}{z_j^2 -1} D_j \nonumber\\
&& + \beta \sum_{j = 1}^N \left [\frac{z_j +z_{j+1}}{z_j -
z_{j+1}}\right] (D_j - D_{j+1}) + \beta \sum_{j = 1}^N \left
[\frac{z_j z_{j+1}+1}{z_j z_{j+1}-1}\right] (D_j + D_{j+1}) \, .
\end{eqnarray}
Here $D_j \equiv z_j \frac{\partial}{\partial z_j}$ while
$\epsilon - \epsilon_0 = (E- E_0) \frac{L^2}{2\pi^2}$. It is worth
pointing out that the above eqation is structurally similar to
that of the $BC_N$ of SM, except $z_{j+1}$ is replaced by $z_{k}$
in our case.

Note that apart from the cyclic symmetry, the Hamiltonian and
hence the $\phi$ equation is also invariant under $z_j \rightarrow
z_j^{-1}$. As a consequence, as in the $BC_N$ Sutherland model
\cite{ser,ohta}, it turns out that even in our case the polynomial
eigenfunctions of $H_1$ with $BC_N$ symmetry as given by Eq.
(\ref{4.5a}) are symmetric polynomials in $(z_j +\frac{1}{z_j})$
i.e. in $\cos(2\pi x_j /L)$.

Finally, let us discuss the solutions to the $\phi$ equation. So
far we have been able to obtain only one solution in the $BC_N$
case but are able to obtain several solutions in the $D_N$ case
and also a few in the $B_N$ and $C_N$ cases.

\subsection{Exact Solution for the $BC_N$ Model}

It is easily checked that the exact solution is
\begin{eqnarray}\label{4.6}
\phi \equiv \phi_{BC_N} = \phi_1 + \alpha \, , \ \epsilon -
\epsilon_0 = 1 + \gamma + 2\gamma_1 + 4\beta \, ,
\end{eqnarray}
where
\begin{eqnarray}\label{4.7}
\phi_1 = \sum_{j=1}^{N} (z_j + \frac{1}{z_j}) \, , \ \alpha =
\frac{2 N\gamma}{1 + \gamma + 2\gamma_1 + 4\beta} \, .
\end{eqnarray}
We will see that this solution (and in fact other solutions, if
any, in the $BC_N$ case) will play important roles in our
construction of solutions for other systems like $B_N, C_N$ and
$D_N$.

\subsection{Exact Solutions For the $B_N$ Model ($\gamma_1 =0$)}

Apart from the obvious solution $\phi (z_j; \beta, \gamma,
\gamma_1 =0)$ as given by Eq. (\ref{4.6}) it turns out that there
are other solutions corresponding to the spinorial representation
for the $B_N$ model. These are
\begin{eqnarray}\label{4.8}
\phi \equiv \phi^{+} = \Pi_{j=1}^{N} (\sqrt{z_j} +
\frac{1}{\sqrt{z_j}}) \, , \ \epsilon^{+} - \epsilon_0 =
\frac{N}{4} [1 + 2\gamma + 4\beta] \, .
\end{eqnarray}

In order to obtain the other solution, we start with the ansatz
\begin{eqnarray}\label{4.9}
\phi = \phi^{+} \psi^{+}
\end{eqnarray}
where $\phi^{+}$ is as given by Eq. (\ref{4.8}) and consider the
equation $H_1 \phi = (\epsilon - \epsilon_0) \phi$. It is easily
seen that in that case $\psi^{+}$ satisfies
\begin{eqnarray}\label{4.10}
H^{+}_1 (z_j; \beta, \gamma, \gamma_1 =0) \psi^{+} = (\epsilon
-\epsilon^{+}) \psi^{+} \, ,
\end{eqnarray}
where $\epsilon^{+}$ is as given by Eq. (\ref{4.8}) while
\begin{eqnarray}\label{4.11}
H^{+}_1 &=& H_1 (z_j; \beta, \gamma, \gamma_1 =0)
+ \sum_{j=1}^{N} \frac{z_j -1}{z_j + 1} D_j \nonumber \\
        &=& H_1 (z_j; \beta, \gamma -1, \gamma_1 =1) \, .
\end{eqnarray}
We thus see that $\psi^{+}$ essentially satisfies the same
equation as satisfied by the $BC_N$ Hamiltonian but with the value
of $\gamma$ and $\gamma_1$ shifted to $\gamma -1$ and 1
respectively. Thus it follows that once we obtain solutions of the
$BC_N$ problem, all these will give us new solutions of the
spinorial type for the $B_N$ model as given by Eqs. (\ref{4.9}) to
(\ref{4.11}). Unfortunately, so far we have been able to obtain
only one solution in the $BC_N$ case as given by Eqs. (\ref{4.6})
and (\ref{4.7}). Using that solution, it then follows that the new
spinorial solution for the $B_N$ model is as given by Eq.
(\ref{4.9}) with energy
\begin{eqnarray}\label{4.9a}
\ \epsilon -\epsilon_0 = 2 + \gamma + 4\beta +\frac{N}{4}[1 +
2\gamma + 4\beta] \, ,
\end{eqnarray}
while $\psi^{+} \equiv \phi_{BC_N} (z_j; \beta, \gamma - 1,
\gamma_1 = 1)$ with $\phi_{BC_N}$ being given by Eqs. (\ref{4.6})
and (\ref{4.7}).

\subsection{Exact Solutions For the $D_N$ Model ($\gamma = \gamma_1 =0$)}

Apart from the above solutions (\ref{4.6}), (\ref{4.8}) and
(\ref{4.9}) (with $\gamma = \gamma_1 =0$), we have found several
other solutions in the $D_N$ case. This is related to the fact
that unlike $B_N$, there are two distinct classes of spinor
representations for $D_N$. Besides, there are also some additional
solutions in this case. It may be noted that in this case the
Hamiltonian $H_1$ acting on $\phi$ is
\begin{eqnarray}\label{4.12}
H_1 (z_j; \beta) = H_1 (z_j; \beta, \gamma =0, \gamma_1 =0) \, .
\end{eqnarray}

The first new solution that we have is given by
\begin{eqnarray}\label{4.13}
\phi \equiv \phi^{-} = \Pi_{j=1}^{N} (\sqrt{z_j}
-\frac{1}{\sqrt{z_j}}) \, , \ \epsilon^{-} -\epsilon_0 =
\frac{N}{4} [1+4\beta] \, .
\end{eqnarray}
Note that the two solutions (\ref{4.8}) (with $\gamma =0$) and
(\ref{4.13}) which correspond to the two different spinorial
representations are degenerate in energy.

In order to obtain the other solution, we start with the ansatz
\begin{eqnarray}\label{4.14}
\phi = \phi^{-} \psi^{-}
\end{eqnarray}
where $\phi^{-}$ is as given by Eq. (\ref{4.13}) and consider the
equation $H_1 \phi = (\epsilon - \epsilon_0) \phi$. It is easily
seen that in that case $\psi^{-}$ satisfies
\begin{eqnarray}\label{4.15}
H^{-}_1 (z_j; \beta, \gamma = 0, \gamma_1 =0) \psi^{-} =(\epsilon
-\epsilon^{-}) \psi^{-} \, ,
\end{eqnarray}
where $\epsilon^{-}$ is as given by Eq. (\ref{4.13}) while
\begin{eqnarray}\label{4.15a}
H^{-}_1 & = & H_1 (z_j; \beta, \gamma = 0, \gamma_1 =0)
+ \sum_{j=1}^{N} \frac{z_j +1}{z_j -1} D_j \nonumber \\
        & = & H_1 (z_j; \beta, \gamma =1, \gamma_1 =0) \, .
\end{eqnarray}
We thus see that $\psi^{-}$ essentially satisfies the same
equation as satisfied by the $B_N$ Hamiltonian but with the value
of $\gamma$ being fixed at 1. Using the solution for the $B_N$
case as given by Eqs. (\ref{4.6}) and (\ref{4.7}) (with $\gamma
=1, \gamma_1 =0$) it then follows that the new spinorial solution
for the $D_N$ model is as given by Eq. (\ref{4.14}) with energy
\begin{eqnarray}\label{4.16}
\ \epsilon -\epsilon_0 = 2 +4\beta +\frac{N}{4}[1+4\beta] \, ,
\end{eqnarray}
while $\psi^{-} \equiv \phi_{BC_N} (z_j; \beta, \gamma=1, \gamma_1
=0)$ with $\phi_{BC_N}$ being given by Eqs. (\ref{4.6}) and
(\ref{4.7}). Notice that this solution is degenerate in energy
with the solution (\ref{4.9}) (with $\gamma =0$).

In addition, we find that the product of the two ``spinorial
solutions'' is also a solution of the $D_N$ model, i.e.
\begin{eqnarray}\label{4.17}
\phi \equiv \phi^{+} \phi^{-} = \Pi_{j=1}^{N} (z_j -
\frac{1}{z_j}) \, , \ \epsilon^{+-} - \epsilon_0 = N[1+2\beta] \,
.
\end{eqnarray}

In order to obtain another solution, as above we start with the
ansatz
\begin{eqnarray}\label{4.18}
\phi = \phi^{+} \phi^{-} \psi^{+-}
\end{eqnarray}
where $\phi^{+}, \phi^{-}$ are as given by Eqs. (\ref{4.8}) and
(\ref{4.12}) respectively and consider the equation $H_1 \phi =
(\epsilon - \epsilon_0) \phi$. It is easily seen that in that case
$\psi^{+-}$ satisfies
\begin{eqnarray}\label{4.19}
H^{+-}_1 (z_j; \beta, \gamma = 0, \gamma_1 =0) \psi^{+-}
=(\epsilon -\epsilon^{+-}) \psi^{+-} \, ,
\end{eqnarray}
where $\epsilon^{+-}$ is as given by Eq. (\ref{4.17}) while
\begin{eqnarray}\label{4.20}
H^{+-}_1 & = & H_1 (z_j; \beta, \gamma = 0, \gamma_1 =0)
+ 2\sum_{j=1}^{N} \frac{z^2_j +1}{z^2_j -1} D_j \nonumber \\
        & = & H_1 (z_j; \beta, \gamma =0, \gamma_1 =1) \, .
\end{eqnarray}
We thus see that $\psi^{+-}$ essentially satisfies the same
equation as satisfied by the $C_N$ Hamiltonian but with the value
of $\gamma_1$ being fixed at 1. Using the solution for the $BC_N$
case as given by Eqs. (\ref{4.6}) and (\ref{4.7}) (with $\gamma =
0, \gamma_1 =1$), it then follows that the new solution for the
$D_N$ model is as given by Eq. (\ref{4.18}) with energy
\begin{eqnarray}\label{4.21}
\ \epsilon -\epsilon_0 = 3 +4\beta +N[1+2\beta] \, ,
\end{eqnarray}
while $\psi^{+-} \equiv \phi_{BC_N} (z_j; \beta, \gamma=0,
\gamma_1 =1)$ with $\phi_{BC_N}$ being given by eqs. (\ref{4.6})
and (\ref{4.7}).

\subsection{Exact Solution For the $C_{N=4}$ Model ($\gamma =0$)}

So far we have discussed all the solutions which are valid for any
N. In addition, in the special case of $N=4$, we have been able to
obtain a solution in the $C_{N=4}$ (and two solutions in the
$D_{N=4}$) case. The solution for the $C_{N=4}$ case is given by
\begin{eqnarray}\label{4.22}
\phi \equiv \phi^{31} = A\bigg
[(z_1+\frac{1}{z_1})(z_2+\frac{1}{z_2})(z_3+\frac{1}{z_3}) + C.P.]
+ B \phi_1 \, ,
\end{eqnarray}
where
\begin{eqnarray}\label{4.23}
\epsilon - \epsilon_0 = 3+6\gamma_1+8\beta \, , \ B =
\frac{8A\beta}{1+2\gamma_1+2\beta} \, .
\end{eqnarray}
Here, by C.P. one means cyclic permutations and $\phi_1$ is as
given by Eq. (\ref{4.6}).

\subsection{Exact Solutions For the $D_{N=4}$ Model ($\gamma = \gamma_1 
=0$)}

Clearly, one obvious solution in the $D_{N=4}$ case is obtained
from the solution (\ref{4.22}) by putting $\gamma_1 =0$. The other
solution is obtained by making use of the ansatz as given by Eq.
(\ref{4.18}), i.e. let
\begin{eqnarray}
\phi = \phi^{31} \phi^{+-} \, ,
\end{eqnarray}
with $\phi^{31}$ being given by Eq. (\ref{4.22}). Using Eqs.
(\ref{4.18}) to (\ref{4.20}) it then follows that this is a
solution for the $D_{N=4}$ model with
\begin{eqnarray}
\epsilon - \epsilon_0 = 13 +16\beta \, , \ B =
\frac{8A\beta}{3+2\beta} \, .
\end{eqnarray}

\section{Conclusions}
In conclusion, we have carried out a systematic study of the
many-body Hamiltonian, related to the short range Dyson model. The
scattering state of this model is obtained and is shown to be a
coherent state. Akin to the CSM, the connection of the scattering
Hamiltonian to a free system is established. Unlike the Calogero
model, analysis of the scattering process for the present model
reveals that, only a part of the partial waves acquire energy
independent phase shifts. We then showed the mapping of the bound
system to decoupled oscillators and by explicitly computing some
of the bound-state eigenfunctions, we find that, the present model
has less degeneracy as compared to the Calogero-Sutherland model.
Finally, we have studied the $A_{N-1}$, $BC_{N}$, $B_N$, $C_N$ and
$D_N$ models on a circle and obtained a part of their excitation
spectrum. A number of open problems, like finding the
$P_{m,q}(x)$'s explicitly for the N-body problem, characterizing
the degeneracy structure, finding the complete eigenspectra and
conserved quantities for these type of models still remains to be
tackled. We hope to come back to some of these issues in future.

M.E, N.G and P.K.P would like to thank Prof. V. Srinivasan and
Prof. S. Chaturvedi for useful discussions.

\end{document}